\documentstyle[epsfig, twocolumn]{esapub}

\def\p{\partial}
\def\apj{ApJ}
\def\aap{A\&A}

\begin{document}

\setlength{\parindent}{0pt}
\setlength{\parskip}{ 10pt plus 1pt minus 1pt}
\setlength{\hoffset}{-1.5truecm}
\setlength{\textwidth}{ 17.1truecm }
\setlength{\columnsep}{1truecm }
\setlength{\columnseprule}{0pt}
\setlength{\headheight}{12pt}
\setlength{\headsep}{20pt}
\pagestyle{esapubheadings}

\title{\bf MACROSCOPIC PROCESSES IN THE SOLAR INTERIOR}

\author{{\bf A.S. Brun$^{1,2}$, S. Turck-Chi\`eze$^1$, J.P. Zahn$^2$} \vspace{2mm} \\
$^1$CEA/DSM/DAPNIA/Service d'Astrophysique, CE Saclay, 91191 Gif-sur-Yvette Cedex, France \\
$^2$D\'epartement d'Astrophysique Stellaire et Galactique, Observatoire de Paris, Section Meudon, \\ 92195 Meudon, France}

\maketitle

\begin{abstract}

With the recent results of helioseismology aboard SOHO, solar models are more and more constrained (Brun, Turck-Chi\`eze and Morel 1998). New physical processes, mainly connected to macroscopic motions, must be introduced to understand these new observations. In this poster, we present solar models including a turbulent pressure in the outer layers and mixing due to the tachocline (Spiegel and Zahn 1992).\\
Our results lead us to conclude that:\\
- Turbulent pressure improves the absolute value of the acoustic mode frequencies ($\sim 10\mu Hz$ at 4 mHz)\\
- Mixing in a tachocline thickness of $0.05 \pm 0.03 R_{\odot}$ (Corbard et al. 1997) looks promising.


  Key~words: sun: interior; sun: turbulent processes.

\end{abstract}

\section{THE SOLAR TACHOCLINE}

\subsection{The Physical Description}

For most purposes, the Sun can be assumed in hydrostatic and thermal equilibrium, neglecting the effects of rotation and magnetic field. However models built with these simplifying assumptions do not agree with the helioseismic data, in particular with  those obtained by the satellite SOHO, and it appears that macroscopic mixing processes must be taken into account not only in the convection zone, but also in the radiative interior. 
Turbulent mixing may be introduced in the models by adding a turbulent term $D_T$ in the diffusive part of the equation for the time evolution of the chemical abundance $X_i$. The equation becomes:
\begin{equation}
\frac{\p X_i}{\p t}=\frac{\p 4\pi\rho r^2 X_i V_i}{\p m} +\mbox{nuclear terms,}\\
\end{equation} 
where the velocity V$_i$ of species i with respect to the center of mass is:
\begin{equation}
V_i=-4 \pi \rho r^2(D_i+D_T)\frac{\p \ln X_i}{\p m}+v_i.
\end{equation}
The velocity $V_i$ is the sum of one term which depends on the concentration gradient $D_i$, and one which does not, $v_i$ (Proffit and Michaud 1991).
In this section, we address the mixing which may occur in the shear layer connecting the differential rotation in the convection zone with the solid rotation of the radiative interior.
Spiegel and Zahn (1992) have given a physical interpretation of this tachocline by invoking anisotropic turbulence, with much stronger viscous transport in the horizontal than in the vertical direction. Such turbulence reduces the differential rotation and therefore inhibits the spread of the layer deep inside the radiative zone. The conservation of mass, momentum and entropy is given by:
\begin{equation}
\frac{\partial \rho}{\partial t}+\vec{\nabla}.(\rho \vec{V}) = 0 
\end{equation}
\vspace{-0.5cm}
\begin{eqnarray}
\rho \left(\frac{\partial \vec{V}}{\partial t}+(\vec{V}.\vec{\nabla})\vec{V}+2\vec{\Omega}\wedge\vec{V}+ \vec{\dot{\Omega}}\wedge\vec{r}\right)  \nonumber \\
 = -\vec{\nabla}P -\rho\vec{\nabla}\Phi+\vec{\nabla}.\parallel\tau\parallel 
\end{eqnarray}
\begin{equation}
\rho T \left(\frac{\partial}{\partial t}+\vec{V}.\vec{\nabla} \right)S = \vec{\nabla}.(\chi \vec{\nabla}T)
\end{equation}
where $\rho$ is the density, $P$ the pressure, $T$ the temperature, $\vec{V}=(u,v,r\hat{\Omega}\sin\theta)$ is the local velocity in the rotating frame, $\hat{\Omega}$ the differential rotation, $S$ the specific entropy, $\parallel\tau\parallel$ the viscous stress tensor, $\Phi$ gravitational potential.\\
In order to solve the system of equations, some simplifying assumptions are made:

\begin{itemize}
\item the flow field is axisymmetric $\vec{V}=\vec{V}(r,\theta,t)$
\item anelastic approximation: $\p \rho/\p t$ is negligible
\item advection is small compared to Coriolis acceleration
\item the tachocline is thin compared to the pressure scale height
\item viscous forces are small compared to Coriolis force.
\end{itemize} 

After separating each variable into its mean value on the sphere plus a perturbation, as $T(r,t)+\hat{T}(r,\theta,t)$, the linearized form of the set of equations, in the stationnary state ($\p/\p t=0$) is (Spiegel and Zahn 1992):
\begin{eqnarray}
\frac{\hat{P}}{P}=\frac{\hat{\rho}}{\rho}+\frac{\hat{T}}{T} \mbox{  equation of state} \nonumber \\
\frac{1}{\rho}\frac{\p \hat{P}}{\p r}=g\frac{\hat{T}}{T} \mbox{  hydrostatic equilibrium} \nonumber \\
-2\Omega r x \hat{\Omega}=\frac{1}{\rho r}\frac{\p \hat{P}}{\p x} \mbox{  geostrophic balance} \nonumber \\
2 \Omega x \frac{\p \Psi}{\p r}=\rho \frac{\p}{\p x}\left[\nu_H(1-x^2) \frac{\p \hat{\Omega}}{\p x}\right]  \nonumber \\
\mbox{  diffusion and advection of angular momentum} \nonumber \\
\frac{N^2}{g}\frac{T}{\rho r^2}\frac{\p \Psi}{\p x}=\frac{1}{\rho c_p r^2}\left[\chi r^2 \frac{\p \hat{T}}{\p r}\right] \nonumber \\
\mbox{  diffusion and advection of heat} \nonumber
\end{eqnarray}

in spherical coordinates ($r$, $x = \cos\theta$), where $\nu_H$ is the horizontal turbulent viscosity ($\nu_H >> \nu_V$) and defining a stream function $\Psi$ for the meridional flows  by:
\begin{equation}
r^2 \rho u=\frac{\p \Psi}{\p x} \mbox{ , }r \rho \sin \theta v=\frac{\p \Psi}{\p r}
\end{equation}

We now project on horizontal eigenfunctions, the variables of the set of equations describing the tachocline:
\begin{eqnarray}
(\hat{P},\hat{T},u)&=&\sum_n (\tilde{P_n},\tilde{T_n},u_n)F_n(x) \nonumber \\
\Psi&=&\sum_n \tilde{\Psi} \int F_n(x)dx \nonumber \\
x\hat{\Omega}&=&\sum_n \tilde{\Omega}_n \frac{d F_n}{dx} \nonumber
\end{eqnarray}
we find that the dominant term for the even eigenfunction is $n=4$.
The final step is to connect the meridional velocity to the differential rotation. Approximating the stream function by $u_4=\tilde{\Psi}_4/\rho r^2$ and using the conservation of angular momentum, we get:
\begin{equation}
\frac{\p \tilde{\Psi}_4}{\p r}=0.5 \nu_H \rho (\mu_4)^4 \frac{\tilde{\Omega_4}}{\Omega}
\end{equation}
with $\mu_4=4.933$. By introducing the variables $\zeta=\mu_4 (r_{bcz}-r)/h$ and $h=r_{bcz}(2\Omega/N)^{1/2}(4K/\nu_H)^{1/4}$ the tachocline thickness, where $r_{bcz}$ is the radius, N the Brunt-V\"ais\"al\"a frequency and $K=\chi/\rho c_p$ the radiative diffusivity at the base of the convective zone, we thus obtain this radial dependancy for $u_4$:
\begin{equation}
u_4(r)=0.5 \frac{\nu_H h}{r_{bcz}^2} \mu_4^3 Q_4 \exp(-\zeta) cos(\zeta) .
\label{u4}
\end{equation}
The anisotropic diffusion invoked to stop the spread of the tachocline will also interfer with the advective transport of chemicals. Chaboyer and Zahn (1992) have shown that the result is a diffusive transport in the vertical direction, with an effective diffusivity given by:
\begin{equation}
D_{T}=\frac{r^2}{D_H}\sum_n \frac{U^2_n(r)}{n(n+1)(2n+1)} 
\end{equation}
where $U_n$ are the coefficients of the expansion of the vertical component of the velocity $u$ in Legendre polynomials.
The eigenfunctions $F_n$ above may be projected on these Legendre polynomials, and to a good approximation one has
\begin{equation}
D_T=\frac{r^2}{D_H} \left(\frac{8}{3}\right)^2 \frac{u^2_4(r)}{180} .
\end{equation} 
Replacing $u_4$ by (\ref{u4}), one reaches the following expression for the vertical diffusivity
\begin{equation}
D_T(\zeta)=\frac{1}{180}\frac{1}{4}\left(\frac{8}{3}\right)^2 \nu_H \left(\frac{h}{r_{bcz}}\right)^2 \mu_4^6 \, Q_4^2 \exp(-2\zeta) cos^2(\zeta).
\end{equation}
We shall treat the tachocline thickness $h$ as an adjustable parameter, roughly bounded between 0.03 and 0.08 $R_{\odot}$ (Corbard et al. 1997).
With the new latitudinal dependence of $\Omega$ (Thompson et al. 1996), $\Omega_{bcz}/2\pi=456-72x^2-42x^4$ nHz, the coefficient $Q_4=1.707\times10^{-2}$ and $\Omega/\Omega_0=0.9104$.

\subsection{Results}
Starting from the reference model of Brun, Turck-Chi\`eze and Morel (1998) built 
with the stellar evolution code CESAM (Morel 1997), which include the recent OPAL 
opacities and EoS, Alderberger et al. (1998) nuclear reaction rates and 
microscopic diffusion (see also Turck-Chi\`eze et al. these proceeding), we 
introduce the coefficient $D_T$ in our diffusion equation with different values 
of $h$ and $N$ (fig. 1). The Brunt-V\"ais\"al\"a frequency $N$ is taken as constant and $h$ is chosen 
such as to yield a tachocline of $\approx 0.05 R_\odot$, as determined by 
(Corbard et al. 1997). In Table 1 and figures 2,3,4,5 we present models 
including the coefficients shown in fig. 1. For one coefficient ($D_{T1}$) we 
also test the influence on our results of the calibration of $Z/X$.

\begin{table*}[!ht]
\begin{center}
  \caption[]{\label{table 4}     
Solar Models}     			
\end{center}
\vspace{-0.6cm}
{\small Parameters description: 
$h$: tachocline thickness, $N$: Brunt-V\"ais\"al\"a frequency, $\alpha$: mixing length parameter, Y$_0$, Z$_0$, (Z/X)$_0$:
initial helium, heavy element and ratio heavy element on hydrogen, 
Y$_s$, Z$_S$, $(Z/X)_S$: idem for surface compositions,  $\tau_b$ 
is the optical depth of the bottom of the atmosphere, R$_{bzc}$, 
T$_{bcz}$ are the radius and temperature at the base of the convective zone,
Y$_c$, Z$_c$, T$_c$, $\rho_c$: central helium, heavy element contents, 
central temperature and density. OPAL/A means that we use OPAL96 opacities and Alexander (1994) for low temperature.}
\begin{center}
   \begin{tabular}{p{3cm}*{6}{c}}  	
    \hline
     Parameters	 &  reference &  $D_{T1}$ &  $D_{T1}$ &  $D_{T2}$ &  $D_{T3}$ \\
    
    \hline
     Opacities   & OPAL/A & OPAL/A & OPAL/A & OPAL/A & OPAL/A\\
     Diffusion    & yes &  yes & yes & yes & yes  \\
     Age (Gyr) & 4.6 & 4.6 & 4.6 & 4.6 & 4.6 \\
     $h$ ($r/R_{\odot}$)& - & 0.05 & 0.05 & 0.08 & 0.05 \\
     $N$ ($\mu$Hz) & - & 100 & 100 & 100 & 10  \\
     (Z/X)$_s$ & fixed & fixed & free & fixed & fixed \\
     \\
     $\alpha$  & 1.768 & 1.754 & 1.763 & 1.748 & 1.759 \\  
     Y$_0$  & 0.2721 &  0.270 & 0.2721 & 0.2693 & 0.2692 \\
     Z$_0$  & 1.96 $10^{-2}$&  1.913 $10^{-2}$ &  1.96 $10^{-2}$ &  1.904 $10^{-2}$ &  1.898 $10^{-2}$\\
     (Z/X)$_0$ & 0.0277 & 0.0269 & 0.0277 & 0.0267 & 0.0267 \\
     \\
     Y$_s$  & 0.2425 &  0.2454 & 0.2475 & 0.2458 & 0.247 \\
     Z$_s$  & 1.803 $10^{-2}$ & 1.796 $10^{-2}$ & 1.841 $10^{-2}$ & 1.79 $10^{-2}$ & 1.795 $10^{-2}$\\
     (Z/X)$_s$ & 0.0244 & 0.0244 & 0.0251 & 0.0244 & 0.0244\\
     $\tau_b$  &2 &  2 & 2 & 2 & 2 \\
     \\
     R$_{bzc}$/R$_{\odot}$ & 0.713 &  0.7145 & 0.7134 & 0.7145 & 0.715 \\
     T$_{bzc} \times 10^6$ (K) & 2.192 & 2.181 & 2.196 & 2.181 & 2.178\\
     \\
     Y$_c$   & 0.642 &  0.639 & 0.642 & 0.638 & 0.641 \\
     Z$_c$   & 2.095 $10^{-2}$& 2.046 $10^{-2}$ & 2.096 $10^{-2}$ & 2.035 $10^{-2}$ & 2.03 $10^{-2}$\\
     T$_c \times 10^6$ (K) & 15.71 &  15.68 & 15.71 & 15.66 & 15.67 \\
     $\rho_c$ (g/cm$^3$) & 153.5 & 153.3 & 153.5 & 153.0 & 153.1 \\
    \hline
   \end{tabular}
  \vspace{-0.3cm}
  \end{center}
\end{table*}
 
\begin{figure}[!ht]
\setlength{\unitlength}{1.0cm}
\begin{picture}(8,5)
\includegraphics{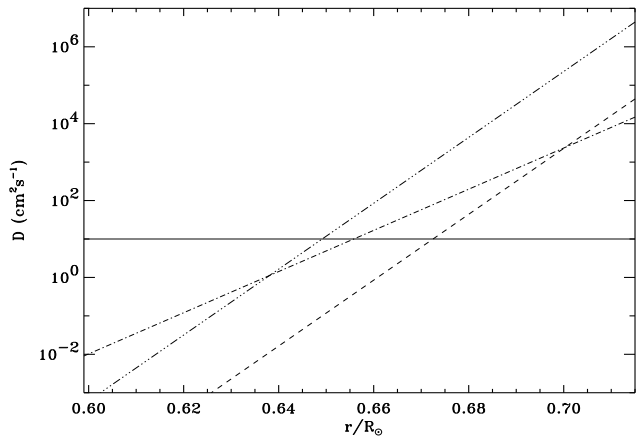}
\end{picture}
\vspace{0.2cm}
\caption{\label{f00} \em  Diffusive coefficients: Approximate value of the microscopic coefficient of $^4He$ (full line) , turbulent coefficient $D_{T1}$ (- - -), turbulent coefficient $D_{T2}$ (dot dash line) and turbulent coefficient $D_{T3}$ (dot dot dot dash line). }
\end{figure}

 Our standard model has a surface abundance of helium of 0.2425 in mass (see 
Table 1), which is a bit too low if we compare with the Basu and Antia (1995) 
value for the OPAL equation of state (Rogers, Swenson and Iglesias 1996), 
$Y_s=0.249 \pm 0.003$.
When allowing for turbulent diffusion in the tachocline, the settling of 
helium is reduced by 16\% for $D_{T1}$ up to 25\% with $D_{T3}$, leading to 
surface abundances $Y_s=0.2454$ or $Y_s=0.247$ which are in better agreement 
with the helioseismic value. Note the smooth composition profile in fig. 3, with its extended plateau below the convection zone.

\begin{figure}[!ht]
\setlength{\unitlength}{1.0cm}
\begin{picture}(8,5)
\includegraphics{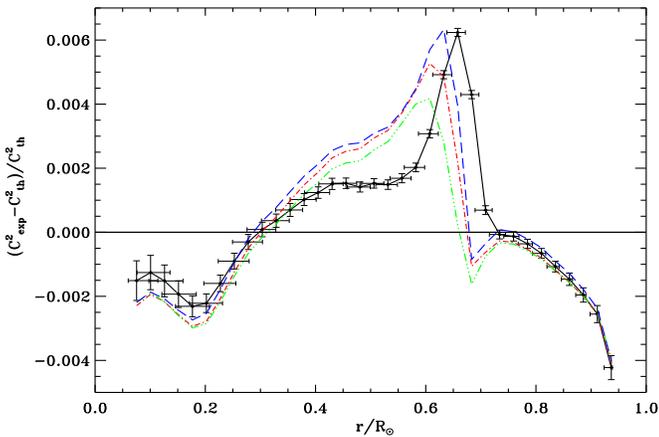}
\end{picture}
\vspace{0.2cm}
\caption{\label{f0} \em Sound Square Speed Difference between GOLF+MDI data, the reference model (full line) and three models calibrated in $Z/X=0.0245$ including a turbulent term: $D_{T1}$ (- - -), $D_{T2}$ (dot dash line) and $D_{T3}$ (dot dot dot dash line). }
\end{figure}

\begin{figure}[!ht]
\setlength{\unitlength}{1.0cm}
\begin{picture}(8,5)
\includegraphics{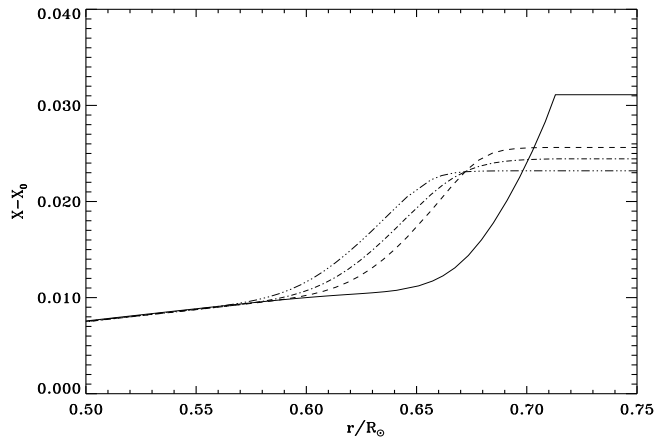}
\end{picture}
\vspace{0.2cm}
 \caption{\label{f3} \em Hydrogen Composition for the reference model (full line) and three models with turbulent mixing due to the tachocline $D_{T1}$ (- - -), $D_{T2}$ (dot dash line) and $D_{T3}$ (dot dot dot dash line). }
\end{figure}

 We remark that the present Z/X value for a mixed model plays an important role 
in this study. If we do not calibrate the Z/X value of the mixed model, starting 
with the heavy element composition $Z_0$ of the reference model, we obtain 
different surface abundances (see for example column 2 and 3 of table 1 for the 
coeficient $D_{T1}$). Also, when comparing the two sound speed profiles (figs. 4 
and 5), we see that in both cases the bump around 0.7 $R_{\odot}$ is pratically 
erased by the introduction of the tachocline mixing, but the calibrated model is 
modified along the whole structure and not only close to the bump. This could be 
understood by the fact that the tachocline mixing inhibits the elements settling 
and then less helium reaches the center, modifying the central conditions.
The remaining part of the discrepancy between the nuclear core and the 
convection zone may come from uncertainties on opacities coefficients or 
knwoledge of the element abundances (See Brun, Turck-Chi\`eze and Morel 1998) or 
from some other mixing. The low value of the tachocline thickness ($h=0.02$) 
inferred by Elliot et al. (these proceedings) could be due to the turbulent coefficient used and also to the fact that they did not calibrate their model in Z/X.

\begin{figure}[!ht]
\setlength{\unitlength}{1.0cm}
\begin{picture}(8,4.8)
\includegraphics{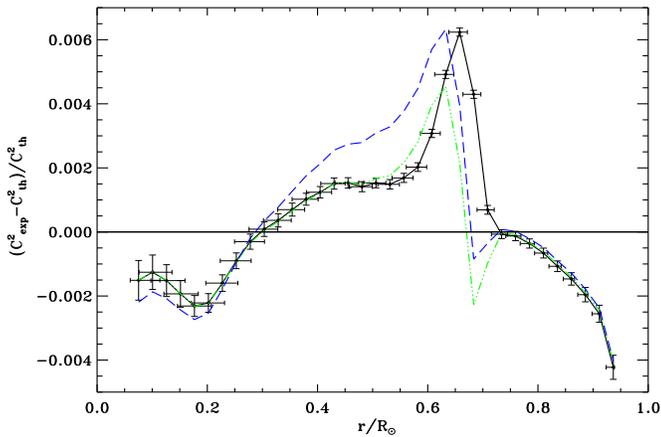}
\end{picture}
\vspace{0.2cm}
\caption{\label{f1} \em Sound Square Speed Difference between GOLF+MDI data, the reference model (full line) and two models including the coefficient $D_{T1}$: calibrated in $Z/X$ (- - -) and with a non calibrated $Z_0=Z_0^{std}=0.0196$ (dot dot dot dash line). }
\end{figure}

\begin{figure}[!ht]
\setlength{\unitlength}{1.0cm}
\begin{picture}(8,4.8)
\includegraphics{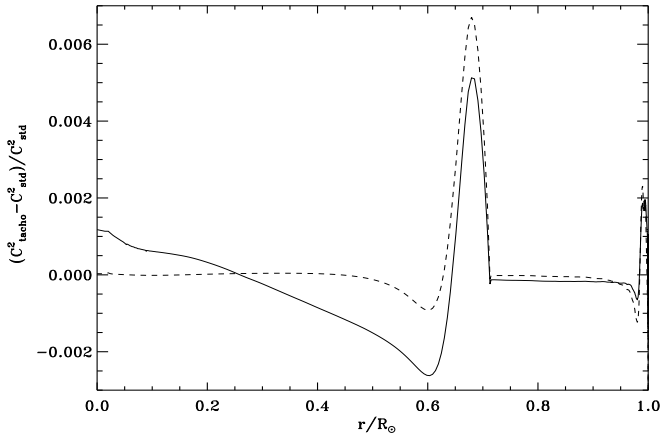}
\end{picture}
\vspace{0.2cm}
\caption{\label{f2} \em The effect of calibration on the sound square speed difference between the reference model and two models including the coefficient $D_{T1}$: calibrated in $Z/X$ (full line) and with a non calibrated $Z_0=Z_0^{std}=0.0196$ (- - -). }
\end{figure}
\vspace{-0.5cm}
\section{EFFECT OF TURBULENT PRESSURE ON THE ACOUSTIC MODES}

We also introduce a turbulent pressure in the total pressure (i.e. $P_{tot}=P_{gas}+P_{turb}$, where $P_{gas}$ is the sum of the gas+radiation+coulombian pressure). 
\vspace{-0.2cm}
\subsection{Implementation of the Turbulent Pressure}
The turbulent pressure is neglected in most solar models, in spite of the fact 
that it represents a substantial fraction of the total pressure in the uppermost  part of the convection zone, which has an important weight in the determination 
of the frequency of the acoustic modes. By definition, the turbulent pressure is given by the horizontal average
 $P_{turb}=<\rho v_z^2>$, $v_z$ being the vertical turbulent velocity,
but here we link it to the convective velocity
$v_{conv}$ drawn from the  Mixing Lenth Treatment (MLT):  
$P_{turb}=\beta \rho \bar{v}_{conv}^2$, where $\beta$ takes into account the 
geometry of the turbulent motions.
We follow Cox and Giuli (1968) who introduce a modified adiabatic gradient 
. 
\begin{equation} 
\nabla ^{'} _{ad} \equiv \nabla _{ad}\left(\frac{dlnP _{gaz}}{dlnP}\right) 
\end{equation}
where $\nabla _{ad}$ is the adiabatic temperature gradient.
Doing so we obtain for the convective efficiency $\Gamma$ and for the velocity $\bar{v}_{conv}$ a corrected expression:
\begin{equation}
\Gamma=\frac{\nabla-\nabla^{'}}{\nabla^{'}-\nabla^{'}_{ad}} \mbox{ , } \bar{v}_{conv}^{2}=\frac{g \delta \Lambda^{2}}{8 \lambda_{p}}\frac{\Gamma}{\Gamma +1}(\nabla-\nabla^{'}_{ad})
\end{equation}
where, $\nabla$ is the temperature gradient of the medium,  $\nabla'$ is the temperature gradient of the moving element, $\Lambda$ is the mixing length, $\lambda_{p}$ the pressure scale height, g the gravitational acceleration and $\delta=-(\p \ln \rho/ \p \ln T)_P$.
After some algebraic manipulations, we deduce a new cubic equation for $\Gamma$:
\begin{equation} \phi_{0}\Gamma^{3}+\Gamma^{2}+\Gamma-A^{2}[\nabla_{rad}-\nabla^{'}_{ad}]=0
\end{equation}
where $\nabla_{rad}$ is the gradient if all the energy is carried by radiation, $A=c_p \kappa g \delta^{1/2} \rho^{5/2} \Lambda^2 / 12 \sqrt{2} a c T^3 P^{1/2}$, $\phi_{0}=9/4$, $c_p$ the specific heat at constant pressure, a the radiation constant, c the light speed and $\kappa$ the opacity coefficient. The real gradient $\nabla$ is thus obtained by:
\begin{equation}
\nabla=\nabla_{ad}^{'}+\frac{\Gamma(\Gamma+1)}{A^{2}}
\end{equation}

We test different values for $\beta$ from 0.5 to 1.5 (see fig. 7), and find that
 the best agreement with direct numerical simulations (see Rosenthal et al. 
(these proceedings)) are obtained for $\beta=1$ (see fig. 6). For the other values, the turbulent contribution to the total pressure is either too low ($P_{turb} \sim 8\%$) or 
too high ($P_{turb} \sim 27\%$).

\begin{figure}[!htb]
\setlength{\unitlength}{1.0cm}
\begin{picture}(8,5)
\includegraphics{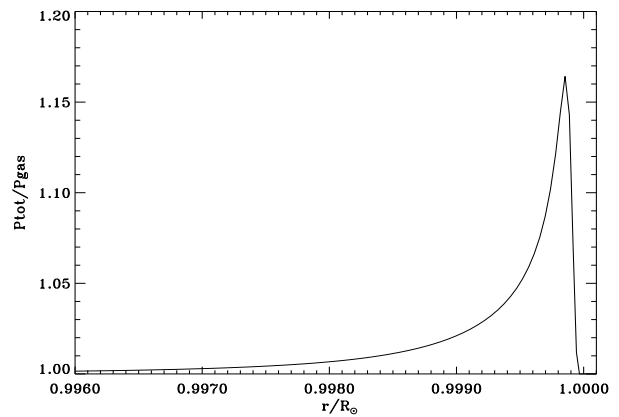}
\end{picture}
\vspace{0.2cm}
 \caption{\label{f5} \em Ratio of the total pressure $P_{tot}=P_{turb}+P_{gas}$ over gas pressure for $\beta=1$. }
\end{figure}
 
\subsection{Results}
By implementing the turbulent pressure in our solar model we obtain the 
following modifications:
\begin{itemize}
\item a higher superadiabatic gradient at its peak near the surface
\item a reduction up to 10 $\mu Hz$ of the l=0 acoustic mode frequencies.
\end{itemize}
This study confirms the important role played by the turbulent pressure in the 
upper layers of the solar convection zone (see Baturin and Mironova and B\"ohmer and R\"udiger these proceedings). Its introduction in the model removes 
a large part of the discrepancy between observed and predicted acoustic frequencies, as 
shown in  fig. 7. Again the best fit is obtained with $\beta=1$. But this 
improvement cannot hide the fundamental shortcomings of the Mixing Length 
Treatment, in particular that of being a local theory which does not allow for 
convective penetration. Serious progress will only be achieved by realistic 
hydrodynamical modelization, as illustrated by the contributions of Rosenthal et
 al. (these proceedings).


\begin{figure}[!htb]
\setlength{\unitlength}{1.0cm}
\begin{picture}(8,5)
\includegraphics{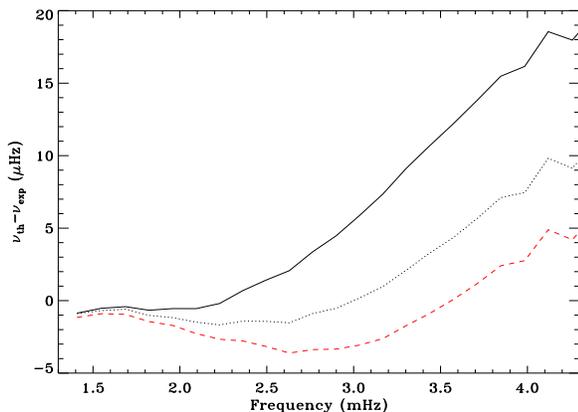}
\end{picture}
\vspace{0.2cm}
 \caption{\label{f6} \em Frequency differences between GOLF l=0 modes and theoretical model with turbulent pressure: $\beta=1.5$ (- - -), $\beta=1$ (dot) and without (full line). }
\end{figure}


\end{document}